\documentclass[%
superscriptaddress,
twocolumn,
 amsmath,amssymb,
 aps,
]{revtex4-1}

\usepackage{graphicx}
\usepackage{dcolumn}
\usepackage{bm}
\usepackage{gensymb}
\usepackage{float}
\usepackage{lineno}
\usepackage{xypic}
\usepackage{ifpdf}
\usepackage{amsmath, amssymb}
\usepackage{mathtools}
\usepackage{qcircuit}
\usepackage{rotating}
\usepackage[utf8]{inputenc}

\newcommand{\ket}[1]{|#1\rangle}

\newcommand{\expec}[1]{\langle #1\rangle}

\begin{document}

\bibliographystyle{apsrev4-1}

\preprint{APS/123-QED}

\title{Finding Broken Gates in Quantum Circuits\--- \\Exploiting Hybrid Machine Learning}

\author{Margarite L. LaBorde}
\affiliation{%
 Hearne Institute for Theoretical Physics and Department of Physics \& Astronomy, Louisiana State University, Baton Rouge, LA 70803\\
}%

\author{Allee C. Rogers}
\affiliation{%
 Hearne Institute for Theoretical Physics and Department of Physics \& Astronomy, Louisiana State University, Baton Rouge, LA 70803\\
}

\author{Jonathan P. Dowling}
\affiliation{Hearne Institute for Theoretical Physics and Department of Physics \& Astronomy, Louisiana State University, Baton Rouge, LA 70803\\
}
\affiliation{National Institute of Information and Communications Technology, Tokyo 184-8795, Japan}
\affiliation{NYU-ECNU Institute of Physics at NYU Shanghai, Shanghai 200062, China}
\affiliation{CAS-Alibaba Quantum Computing Laboratory, USTC, Shanghai 201315, China}


\date{\today}

\begin{abstract}
Current implementations of quantum logic gates can be highly faulty and introduce errors. In order to correct these errors, it is necessary to first identify the faulty gates. We demonstrate a procedure to diagnose where gate faults occur in a circuit by using a hybridized quantum-and-classical K-Nearest-Neighbors (KNN) machine-learning technique. We accomplish this task using a diagnostic circuit and selected input qubits to obtain the fidelity between a set of output states and reference states. The outcomes of the circuit can then be stored to be used for a classical KNN algorithm. We numerically demonstrate an ability to locate a faulty gate in circuits with over 30 gates and up to nine qubits with over 90\% accuracy.
\end{abstract}

\pacs{Valid PACS appear here}
\maketitle


\section{\label{sec:level1}Introduction}
Quantum computers are becoming more realizable as we approach the noisy intermediate-scale quantum (NISQ) era \cite{NISQ}. Tools like the IBM Q-Experience allow researchers to program and simulate quantum algorithms on a real quantum computer with a small number of qubits. These quantum computers are programmed using quantum logic gates, which act on the qubits to perform different operations; however, current implementations of these gates are prone to physical faults such as extraneous phase shifts or rotations, which introduce systematic errors into the system \cite{Fault,Models}. Before error correction protocols can be implemented, it is necessary to identify the gate producing the error. Here, we propose a preprocessing step to diagnose gate faults---without altering the circuit itself---by utilizing machine learning.

Machine-learning techniques are powerful tools for classification and pattern recognition, and much work has been done to determine the potential advantages of quantum machine-learning algorithms \cite{Schuld, Torlai, Support}. We consider a hybrid quantum-classical machine learning technique that utilizes both quantum and classical algorithms. Similar hybrid schemes have been used to achieve machine-learning capabilities for NISQ devices \cite{Boolean, Sumeet}. Using a hybrid technique, we harness the computational advantage of quantum systems while utilizing more freely available classical resources such as memory. Here, we consider a machine-learning algorithm known as K-Nearest-Neighbors.

\begin{figure}[h]
\begin{center}
\includegraphics[width=3 in]{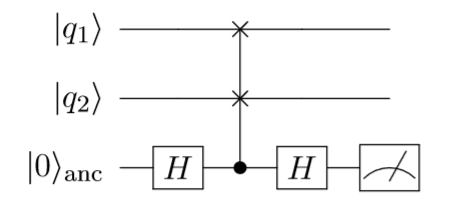} 
\caption{\label{fig:Swap} Swap-test circuit as used in Ref. \cite{SwapTest}. The probability of measuring the ancillary bit to be in the $\ket{0}$ state directly relates to the fidelity between states $\ket{q_0}$ and $\ket{q_1}$. This fidelity provides the Euclidean distance between two quantum states.}
\end{center}
\end{figure}

K-Nearest-Neighbors (KNN) is a comparatively simple classification algorithm. KNN takes a training set of $d$-dimensional vectors that are all labeled with their respective classifications. Given a new unclassified vector, KNN determines the class of the vector from the most common class of the $k$-nearest training vectors. Typically, the Euclidean distance determines the distance measure between vectors. 
In quantum states, the overlap or fidelity between two states acts as a similarity measure that is analogous to the Euclidean distance \cite{Fidppl}, and this fidelity is found through a simple circuit known as a swap test, as shown in Fig. \ref{fig:Swap} \cite{SwapTest}. 
This swap-test circuit can use carefully prepared state vectors to evaluate distances between classical vectors in KNN-style algorithms \cite{Microsoft, Notmicrosoft}. 

We utilize a modified version of the swap-test circuit that acts on multi-qubit states. With this circuit, we define a hybrid quantum-classical machine-learning technique that compares the output state of a quantum circuit to a series of reference states. We then determine, from the output of KNN classification, where in the test circuit a gate fault occurs.  We show that with relatively simple reference states and carefully chosen inputs, we are able to achieve simulated accuracies over 90\%, even for relatively large quantum circuits.

\section{\label{sec:level1}Gate-Fault Classification}

We assume that the circuit under test (the test circuit) is of known composition and has only one physical gate fault. Furthermore, we assume that this gate fault can be modelled by a physical fault, such as an unintentional rotation or additional phase factor. In the case of controlled-NOT (CNOT) gates, we also consider the possibility of a misplaced target or control qubits, such that the intended operation is implemented incorrectly on a certain qubit rather than another.

We simulate circuits by generating combinations of random unitary gates, Hadamard gates, and CNOT gates. We model random unitaries using the property that any unitary gate can be decomposed into elementary quantum gates, as shown in Ref.\cite{Elementary}. Any unitary operation on $n$ qubits can be decomposed into combinations of one-qubit unitaries of the form:
\begin{equation}
    A = \Phi(\delta)R_z (\alpha) R_y(\theta) R_z(\beta)\,,
\end{equation}
where $R_z$ and $R_y$ are rotations on the Bloch sphere around their respective axes, and $\Phi$ is a phase-shifting gate. The angles $\delta$, $\alpha$, $\theta$, and $\beta$ are the specific parameters that determine the gate $A$. We include Hadamard gates, CNOT gates, and phase-shifting gates along with this combination, which allows us to simulate a general quantum circuit, since they form a set of universal two-qubit gates \cite{Uni3}. Hadamard gates can be decomposed either in a manner similar to the unitary gates, or through the Reck decomposition \cite{Reck}. We simulate physical defects by altering the arguments of the decomposition from 0 to $2\pi$ or by considering permutations of the CNOT gate (that is, a CNOT gate operating on various combinations of target and control qubits other than the intended target and control combination).

\begin{figure}[t]
\begin{center}
\includegraphics[width=3 in]{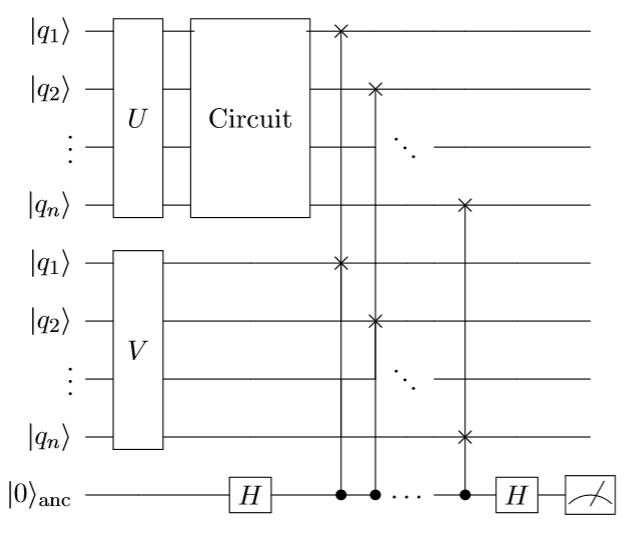} 
\caption{\label{fig:Diagnostic}Diagnostic circuit for the hybrid KNN technique. A string of $n$ qubits in either the $\ket{0}$ or $\ket{1}$ states are fed into unitaries $U$ and $V$. $U$ is controlled to specify the input state, and $V$ is controlled to specify the reference state. The input state is then fed into the test circuit. Afterwards, a series of Fredkin gates perform a controlled SWAP operation between the corresponding qubits of the circuit and the qubits of the reference state. Each SWAP gate is controlled by an ancilliary qubit $\ket{0}_{\rm{anc}}$, which gives a measure of distance between the output of the circuit and the reference state. For input tests, $V$ is set to be the Quantum Fourier Transform (QFT).}
\end{center}
\end{figure}

Once a test circuit is generated, the KNN training data is populated. We receive this training data from the output of a diagnostics circuit, which is a modified form of the swap-test circuit. The diagnostics circuit assumes control of the input to the test circuit, which can be controlled by sending a string of qubits $\ket{q_0}\ket{q_1}...\ket{q_n}$, all of which are in either the $\ket{0}$ or $\ket{1}$ state. The output of $U$ is the input to the circuit. Then the output of the circuit on this state is related to a reference state through many successive controlled-SWAP, or Fredkin, gates. A single ancillary qubit controls these Fredkins gates. The reference state is manipulated by sending the same string of input qubits into a unitary $V$. (See Fig \ref{fig:Diagnostic}.)

When comparing two single qubits, if the output state of the test circuit $\ket{\Psi}$ is exactly the same as the reference state $\ket{\Phi}$, then the probability of measuring the ancillary bit to be in the $\ket{0}$ state is one; however, if the states are slightly different, the probability of measuring the ancilla in the zero state is determined by:

\begin{equation}
\label{eq:probs}
P(\ket{0}_{\rm{anc}}) = \frac{1}{2} + \frac{1}{2}|\expec{\Psi|\Phi}|^2\,,
\end{equation}

We can generalize this for comparisons of multi-qubit states. If the two states are not identical, the probability is given by:

\begin{equation}
P(\ket{0}_{\rm{anc}}) = \frac{1}{2^n} + \frac{1}{2^n}\sum_{i=1}^{n}|\expec{\Psi_i|\Phi_i}|^2 + O(\epsilon),
\end{equation}
where this sum of the respective $i^{\rm{th}}$ qubits of the two states and $n$ is the number of qubits inputted to the test circuit. This probability is approximately the fidelity between the two states when cross-terms between any $i^{\rm{th}}$ and $j^{\rm{th}}$ qubits become vanishingly small.  

\begin{figure*}[t!]
\begin{center}
\includegraphics[width=\textwidth]{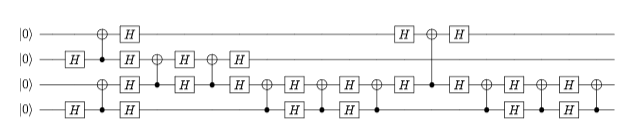} 
\caption{\label{fig:repeater} This quantum repeater circuit \cite{Repeater} is used as a test for the KNN algorithm. Since it is entirely composed of Hadamard and CNOT gates, it should be more difficult to classify than a less repetitive circuit. With unitary-controlled input states, the four-dimensional KNN algorithm achieves accuracies of approximately 95\%. By restricting ourselves to only inputting $\ket{0}$ states, as pictured above, the accuracy drops down to 78\% for the same classification algorithm, or to 80\% with more reference states.}
\end{center}
\end{figure*}

These probabilities are stored classically. Using $d$ different input and reference state pairs, we construct a $d$-dimensional classical vector \textbf{S} that contains the measured probability for each set of states such that:

\begin{equation}
\boldsymbol{S}=[P_1,P_2, .., P_d]\,.
\end{equation}
Here, $P_i$ is the probability, as described in Eq. \ref{eq:probs}, associated with the $i^{th}$ pair of comparison states. This vector is then stored and retrieved classically, allowing for its use in a classical KNN algorithm. This algorithm comes equipped with a set of training vectors that are already classified. We then provide new vectors for the algorithm to classify. The algorithm obtains the Euclidean distance between the new vector and each vector in the set of training data. We specify a parameter, $k$, and the $k$-nearest-neighbors are queried for their assigned class. The input vector obtains the class of the majority of its neighbors. In this case, the classes of the algorithm are the identities of the various gates in the circuit. This procedure can be weighted so that more emphasis is placed on the classes of vectors closer to the input vector, and $k$ can be adjusted to increase accuracy. The diagnostic circuit provides the new input vector, which, in turn, is classified using KNN. Thus the output of the machine-learning algorithm is the identity of the faulty gate.

\section{\label{sec:level1}Results}

We simulated results of the diagnostics algorithm for both randomly generated circuits as well as a known test circuit. For each circuit generated, 200 different random errors were created for each gate. Eighty percent of this data was used to train the classical KNN algorithm. The remaining twenty percent was employed as a test.  Each circuit was queried using four comparison states, which means the vectors used for the KNN protocol were four-dimensional.

\subsection{\label{sec:level2}Quantum Repeater Circuit}
As a preliminary test run, we simulate the results on a known circuit as a proof-of-concept. We use the circuit given in Ref. \cite{Repeater}, also shown in Fig. \ref{fig:repeater}. This protocol was chosen since it uses only four qubits and has approximately thirty gates which are all either CNOT or Hadamard gates. Since this circuit is highly repetitive, the algorithm struggles more to accurately classify which gate is faulty, since it can be easily confused by identical gates at different portions of the circuit. Therefore, this circuit is a rather extreme example.

This circuit also requires a specific input state (namely all $\ket{0}$) to function as designed. As such, we consider both the case where we control the input, as before, and the case where we do not. For this circuit, the average simulated accuracy with input control was around 95\% for both the cases when $V$ was a multi-qubit Hadamard or the QFT. Enforcing the restriction to the all-zero input case, the average accuracy dropped to 78\% when using the same comparison states. 
By altering the number of reference states or the $k$ parameter, it is possible to raise this value slightly to 80\%. 

\subsection{\label{sec:level2}Randomized Circuits}

\begin{figure}[ht]
\begin{center}
\includegraphics[width=3.5 in]{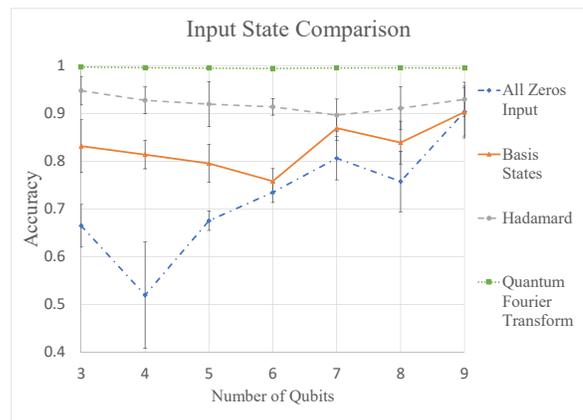} 
\caption{\label{fig:10gates} A comparison of different input states to the test circuit when scaling the number of qubits. The All Zeros Input data refer to an input of only $\ket{0}$ states, the Basis States data refers to an input of a series of $\ket{0}$ and $\ket{1}$ states, and that same series of $\ket{0}$ and $\ket{1}$ is fed either into a multi-qubit Hadamard or a QFT for the final two sets of data. Both the Hadamard and the QFT consistently give accuracies above 90\% for all simulated data.}
\end{center}
\end{figure}

\begin{figure}[ht]
\begin{center}
\includegraphics[width=3.5 in]{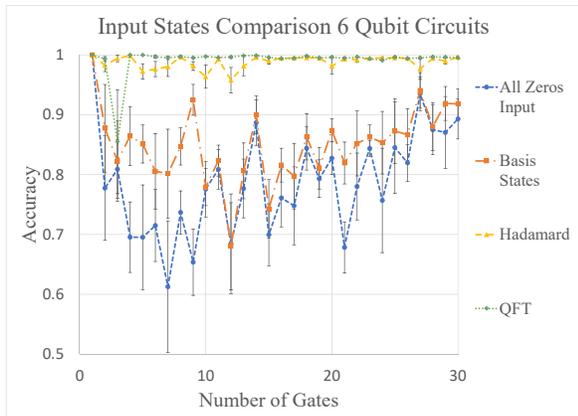} 
\caption{\label{fig:6qubits} A comparison of different input states to the test circuit when scaling the total number of gates. The All Zeros Input data refer to an input of only $\ket{0}$ states, the Basis States data refers to an input of a mix of $\ket{0}$ and $\ket{1}$ states and that same mix of $\ket{0}$ and $\ket{1}$ is fed either into a multi-qubit Hadamard or a QFT for the final two sets of data. Both the multi-qubit Hadamard and the QFT produce high accuracy even for large circuits.}
\end{center}
\end{figure}

We use the results from randomly generated circuits to determine appropriate choices for the unitaries $U$ and $V$ (Fig. \ref{fig:Diagnostic}). Theoretically, the ideal reference state would be the output of the test circuit when no fault has occurred \cite{SwapTest,Microsoft}; however, in practice these outputs could be complicated to produce and may require a working duplicate of the test circuit. We therefore look for comparatively simple unitaries that can be reliably implemented and are stable --- this choice provides a high degree of accuracy for a variety of circuits. Choices for the input and reference states included the following: only using all $\ket{0}$ states in the input (denoted All Zeros Input in the figures), a mix of $\ket{0}$ and $\ket{1}$ states (denoted Basis States), the same mix of $\ket{0}$ and $\ket{1}$ states operated on by a multi-qubit Hadamard gate (Hadamard), and the same mix of zero and one states acted on by a quantum Fourier transform (QFT). The all zeros, basis, and Hadamard options were chosen due to their simplicity and repeatability. The QFT was chosen due to its property of being a maximally mixing unitary. 

Upon simulating the various choices for reference states, the QFT was shown to be the most stable and accurate reference, and thus it is used as the reference unitary $V$ for all input tests. In Fig \ref{fig:10gates}, we compare the various choices for $U$ as the number of qubits in the test circuit increases. Although matching the reference state to the input with a QFT performs optimally, at a steady rate of 99\% accurate, the relative complications in implementing the QFT make that combination less practical. In comparison, using a QFT reference state and letting $U$ be a multi-qubit Hadamard achieves simulated accuracies of greater than 90\% for up to nine qubits\---while being simpler to implement. 

The combination of the Hadamard and QFT unitaries performs similarly well when considering six qubit circuits of various lengths. (See Fig. \ref{fig:6qubits}) Compared to the input states which are not modified by a unitary transformation, the Hadamard and QFT show little dependence on the length of the circuit itself or in the number of qubits. The latter is likely due to the scaling of the diagnostics circuit with respect to the number of qubits, and the former arises from the companion fact that the classical KNN algorithm used to classify faulty gates is kept to a relatively small dimension. These two properties circumvent the loss of accuracy seen in typical KNN schemes when the dimensionality of the training space is allowed to grow.

\section{Discussions and Conclusions}
We propose a hybrid quantum and classical machine learning algorithm capable of identifying the faulty gate in a given circuit. Using a set of unitary gates to control the input and reference states, we show simulated accuracies of greater than 90\% for up to nine qubits and 30 gates in a circuit. In all general cases, we have considered only a four dimensional KNN algorithm. The number of dimensions can be altered for specific implementations when necessary to improve accuracy. In the example repeater circuit, where the input cannot be meaningfully manipulated, using more reference states\---and thus increasing the dimensionality of the classical algorithm\---has shown increased accuracy to around 80\%. 

\section*{Acknowledgements}
We would like to acknowledge the Air Force Office of Scientific Research, the Army Research Office, and the National Science Foundation. The authors would also like to thank Nathan Miller and Lior Cohen for helpful discussions.

\bibliography{apssamp}

\end{document}